\documentclass[twocolumn,aps]{revtex4}
\usepackage{epsfig}
\usepackage{amsmath}

\newcommand{\be}{\begin{equation}}
\newcommand{\ee}{\end{equation}}
\newcommand{\bea}{\begin{eqnarray}}
\newcommand{\eea}{\end{eqnarray}}

\begin{document}
\title{Comments on ``Note on varying speed of light theories''}
\author{Jo\~{a}o Magueijo$^{1,2,3}$ and John W. Moffat$^1$}
\address{$^1$Perimeter Institute for Theoretical Physics, 31
Caroline St N, Waterloo N2L 2Y5, Canada\\
$^2$ Canadian Institute for Theoretical Astrophysics,
60 St George St, Toronto M5S 3H8, Canada\\
$^3$ Theoretical Physics Group, Imperial College, Prince Consort
Road, London SW7 2BZ, England}

\begin{abstract}
In a recent note  Ellis criticizes  varying speed of light
theories on the grounds of a number of foundational issues. His
reflections provide us with an opportunity to clarify some
fundamental matters pertaining to these theories.
\end{abstract}

\pacs{PACS Numbers: *** }
\date{\today}

\maketitle

\section{Generalities}

In a recent publication~\cite{Ellis}, varying speed of light (VSL)
theories were criticized for failing to address a number of
foundational issues~\cite{Uzan}. The criticism allegedly applies to
the entirety of the extensive literature in the field, although the
author says that he ``will not comment on any specific such
papers''. Although an extensive review is cited~\cite{Magueijo}, it
is clear that perusal of the references cited therein was deemed
unnecessary, including early pioneering
papers~\cite{Moffat,Moffat2,Albrecht,Barrow,Magueijo2,Clayton}. We
feel, nonetheless, that Ellis's effort provides an opportunity to
clarify a few foundational matters related to these theories.

Before embarking on an examination of the five points he has
raised, we stress that VSL means by now a number of very distinct
classes of theories, and
one cannot generalize. Broadly speaking, these theories fall into
two categories: those where $c$ varies in space-time, and those
where it varies with the energy scale. The former were motivated
by cosmology, the latter by phenomenological quantum gravity. One
may further categorize VSL theories according to the fashion in
which they contradict (or adapt) Lorentz invariance. As an
illustration we shall consider the following:
\begin{itemize}
\item Theories with soft breaking of Lorentz invariance
(e.g.~\cite{Moffat,Magueijo2});

\item Theories with hard breaking of Lorentz invariance (e.g.~\cite{Albrecht,Barrow});

\item Bimetric theories, for which there is a metric for matter
and another for gravity
(e.g.~\cite{Clayton,Clayton2,Clayton3,Clayton4,Drummond});

\item Non-linear or deformed special relativity,
``DSR'' (e.g.~\cite{Smolin,Smolin2,Kimberly}).
\end{itemize}
This classification is far from extensive (for more examples,
see~\cite{Magueijo}), but has the merit of avoiding the perils of
sweeping generalizations.

One should then avoid getting tangled up in matters of
terminology~\cite{Ellis,Uzan}: of course VSL is an umbrella term
for what could be variously called varying maximal attainable
speed, varying speed of massless particles, varying speed of
particles that follow the null cone, varying $c$ as in the $c$
that appears in many equations in modern physics, varying speed of
photons, etc. People who use the term VSL are well aware of these
distinctions, but they believe in the benefits of a simple
terminology. Which $c$ is it? One should examine the question for
VSL theories on a case by case basis. Pedantic terminology can
hardly be a substitute.

\section{The speed of light and measurement}

Following Ellis~\cite{Ellis}, let us first consider $c$ as the
speed of the photon. Can $c$ vary? Could such a variation be
measured? As correctly pointed out by Ellis, within the current
protocol for measuring time and space the answer is no. The unit
of time is defined by an oscillating system or the frequency of an
atomic transition, and the unit of space is defined in terms of
the distance travelled by light in the unit of time. We therefore
have a situation akin to saying that the speed of light is ``one
light-year per year'', i.e. its constancy has become a tautology
or a definition.

But then, within such a framework, neither can the constancy of the
speed of light be falsified, thus losing its status as a scientific
statement. The constancy of $c$ can only be a scientifically sound
concept if its variability is a possibility. The physical meaning of
a constant $c$ has to lie elsewhere, beyond the convenient but not
necessary definitions of units. Instead of embarking on that search,
Ellis goes on to say that only the variability of dimensionless
constants can be discussed~\footnote{It is true that the results of
calculations in physical theories ultimately have to be expressed as
dimensionless numbers: computers can only process dimensionless
numbers. And, of course, a measurement is always dimensionless,
because it is the ratio of what you are measuring and a unit. But it
is the definition of the unit that sneaks in as a dimensional
statement that appears like pure numbers. With its choice you are
saying that the unit of measurement (which obviously has units) does
not vary.}. This is a misleading statement: what about varying
$G$~\cite{Dirac} and $e$ theories, where $G$ denotes Newton's
gravitational ``constant'' and $e$~\cite{Bekenstein,Barrow2} the
charge of an electron? Can the Hubble ``constant'' $H$ not vary?

An historical analogy may be of use here. Consider the
acceleration of gravity, little $g$. This was thought to be a
constant in Galileo's time. One can almost hear the Ellis of the
day stating that $g$ cannot vary, because ``it has units and can
always be defined to be constant''. The analogy to the present day
relativity postulate that $c$ is an absolute constant is
applicable, for the most common method for measuring time in use
in those days did place the constancy of $g$ on the same footing
as $c$ nowadays. If one insists on defining the unit of time from
the tick of a given pendulum clock, then the acceleration of
gravity is indeed a constant by definition. Just like the modern
speed of light $c$. And yet the Newtonian picture is that the
acceleration of gravity varies, the status of constant being
shifted to $G$, determining the proportionality between $g$ and
$M/r^2$. Why do we do it? Although this would never be phrased in
this way in Newton's time, it is because we insist on naturally
defining the unit of time by means of electromagnetic clocks. The
dynamics tells us that it is a bad idea to define the unit of time
by means of a pendulum clock, for the acceleration of gravity
varies {\it and that is a dimensional statement.} Ultimately, the
dynamics is rendered simpler in units for which $g$ varies. Of
course we could still define $g$ to be a constant, even within the
context of Newtonian (as opposed to Galilean) dynamics but the end
result would be needlessly complicated.

Likewise, a VSL theory is one for which the dynamics is rendered
simpler by choosing units for which $c$ varies. One can always
define units such that $c=1$ (in the same way that one can always
define units for which $g=1$), but the dynamics naturally picks a
set of simpler units, for which $c$ varies. This can be shown by
explicitly taking any of these theories and performing a change of
units leaving $c$ constant (this was done in one
case~\cite{Magueijo}). The exercise is similar to changing units, so
that the Hubble ``constant'' $H$ is indeed a constant (the Friedmann
equations get badly messed up), or the universe is not expanding and
instead we are all contracting (we lose minimal coupling to
gravity). A crucial aspect in VSL theories is the breaking of
Lorentz invariance, as discussed further in the next Section. With
such a glaring fundamental dynamical postulate, it would be
unacceptable to define units of space and time the way they are
currently defined. Ultimately, in theories in which some of the
postulated constants of physics are assumed to vary, it is to be
understood {\it that these physical quantities were never truly
constant} throughout the age of the universe --- they only appear to
be constant to observers during specific epochs.  So it is perfectly
acceptable to consider them as variable quantities, such as the
temperature of air, even though they are not dimensionless
quantities.

How will then a hypothetical future physicist define units, if VSL
were correct? Firstly, in some VSL theories the fine-structure
constant $\alpha$ varies in space and time. Concomitantly all the
atomic transitions used to define the unit of time also vary, so
that blindly using atomic clocks would be as impractical as using a
pendulum clock on a mission to Mars. We could of course still use
the same procedure, based on atomic transitions, but only as long as
{\it we used the varying-$\alpha$ theory itself to correct the
clock}, just like we could use a pendulum clock on the Moon as long
as we used Newton's laws to correct for the lower acceleration. In
other words, you can keep using electromagnetic clocks, but you will
need to use the adopted theory to correct them.

The same will happen for the definition of the unit of space. It can
be defined from the unit of time and the speed of the graviton
(bimetric theories~\cite{Clayton,Clayton2,Clayton3,Clayton4,
Clayton5,Drummond,Visser}) or the speed of low energy photons
(DSR)~\cite{Smolin,Smolin2,Kimberly}. In theories with explicit
breaking of Lorentz invariance, one can use a reference $c_0$,
inferred from the physical (variable) $c$ and possible effects
proportional to $\dot c/c$ (such as violations of energy
conservation) predicted by a particular choice of theory. In other
words, one could correct the physical $c$ in order to infer a
constant $c_0$ precisely using the dynamics of the theory. The $c_0$
could then be used to define the unit of length.

This is just one possible strategy for defining the units
of space and time in a VSL world. It is based on adapting existing
protocols and we suspect there might be more direct alternatives.

\section{The speed of light and the Lorentz group}

A crucial point complementing the discussion of the previous
Section is the issue of the fate of Lorentz symmetry under VSL,
issue 3 in~\cite{Ellis}. Let $c$ now represent what in special
relativity is the invariant speed when transforming between
inertial observers. Only this invariance of special relativity
justifies the definitions of units currently in use. Before 1905
no one would have thought of defining units in the modern way. All
theories under the name of VSL break Lorentz invariance in some
way. For the ether theories in vogue before 1905, it was possible
to define an absolute frame of reference with respect to the ether
in which the speed of light is constant; in all other inertial
frames of reference $c$ would vary.

The question is then: If Lorentz invariance is broken, what
happens to the speed of light? Given that Lorentz invariance
follows from two postulates --- (1) relativity of observers in
inertial frames of reference and (2) constancy of the speed of
light---it is clear that either or both of those principles must
be violated. Thus VSL appears almost inevitably associated with
the breaking of Lorentz invariance.

It is pointed out correctly by Ellis, that all VSL theories should
then make concrete proposals for what happens to Lorentz symmetry.
This is in fact at the core of the definition of any such theory.
Bimetric theories, for example, replace Lorentz symmetry by two
copies of the group
$SO(3,1)$~\cite{Clayton,Clayton2,Clayton3,Clayton4,Clayton5,Drummond,Visser}.
In the most general case, the dynamics is in the $SO(3,1)$
dynamical matrix relating the two tetrads ({\it vierbeins})
representing the group.

In DSR theories the Lorentz group is replaced by a non-linear
representation~\cite{Smolin,Smolin2,Kimberly}. One example is the
set of transformations between 4-momenta: \bea \label{fltransp}
p_0'&=&{\gamma\left(p_0- vp_z \right)
\over 1+l_P (\gamma -1) p_0  -l_P\gamma v p_z }\\
p_z'&=&{\gamma\left(p_z- v p_0\right)
\over 1+l_P (\gamma -1) p_0  -l_P\gamma v p_z }\\
p_x'&=&{p_x
\over 1+l_P (\gamma -1) p_0  -l_P\gamma v p_z }\\
p_y'&=&{p_y \over 1+l_P (\gamma -1) p_0  -l_P\gamma v p_z } \eea
The choice of such non-linear representations renders the
definition of duals highly non-trivial~\cite{Kimberly}. This
has an impact upon the definition of the space-time metric (to be discussed
in the next Section).

In the theories discussed in~\cite{Magueijo2} a distinction is made
between the coordinate $x^0$ (with dimensions of space) and time (as
measured by a minimally coupled clock). The distinction allows for a
varying-$c$ (controlled by a dynamical field), yet it is possible to
preserve most of the features of Lorentz invariance which are
operationally meaningful (such as those associated with the outcome
of the Michelson-Morley experiment). Here the couplings to the
matter action are essential in pegging down as the simplest a system
of units for which $c$ does vary. Local Lorentz invariance can also
be violated either spontaneously~\cite{Moffat,Moffat2,Moffat4} or
dynamically (hard breaking)~\cite{Albrecht}: these are the only
theories where both postulates of relativity are violated.

\section{The speed of light and the metric}
Next comes the issue of general covariance. How does a varying $c$
enter into the metric of a general space-time and how does it
change under a coordinate transformation? It is claimed that a
varying $c$ is necessarily something that can be undone by a
coordinate transformation. However this is not true: The $c$ in
VSL theories is never a coordinate speed of light. It is the
physical speed of light measured by free-falling observers and
cannot be undone by a coordinate transformation.

Nonetheless Ellis raises an important question. Each VSL theory must specify
how diffeomorphism invariance or lack of it works, so that the
statement that the varying $c$ is invariant under coordinate
transformations is true. Again a generalization
is not possible, for each type of VSL theory has a different
answer to this question.

For example in bimetric theories the diffeomorphism
transformations of frames occur without changing the ratio,
$c_\gamma/c_g$, between the speed of the photon and gravity.
In the tetrad formalism of spontaneous
violation of Lorentz invariance, the non-vanishing vacuum
expectation value of a massive vector field $\langle A_\mu\rangle$
also breaks diffeomorphism invariance as well as local Lorentz
invariance~\cite{Moffat,Moffat2,Moffat4}. Therefore, in this
theory the initial speed of light in the early universe $c_i$ and
the currently measured speed of light $c_0$, resulting, say, from
a first order phase transition in $c(t)$ (where $c$ is treated as
an order parameter) cannot be undone by a transformation from $t'$
to $t$ during the period when the spontaneous violation of the
Lorentz group $SO(3,1)$ and the group of diffeomorphism
transformations takes place.
%In a scenario with dynamical symmetry
%breaking of Lorentz invariance, the diffeomorphism invariance may
%not be violated and a different dynamical interpretation of the
%varying speed of light $c$ is realized.

DSR theories, on the contrary, replace the usual metric by the
so-called rainbow metric~\cite{Smolin2}, i.e. the metric runs, and
we have a different metric for each energy scale. This does not
contradict the principle of relativity (and indeed it is implied
by this principle) because a non-linear representation for the
Lorentz group has been adopted. Diffeomorphism transformations
change the metrics without changing the ratio between the speeds
of photons with different frequencies. An explanation of the
structure of these theories, their Einstein equations, and the
impact on solutions such as black hole and cosmological solutions
in presented in~\cite{Smolin2}.

Yet another approach is that adopted in the theories of the type
described in~\cite{Magueijo2}. Here a $x^0$ coordinate, with the
units of length, should be used in the definition of metric and
diffeomorphisms, but it is not to be converted into time in the
usual way, as explained in~\cite{Magueijo2}. The physical $c$
appears in the relation between this coordinate and the physical
time (measured by clocks defined by minimal coupling of a scalar
field to the matter action) and is left invariant by
diffeomorphisms (after the concept is suitably modified).

Finally there are VSL theories  which crudely violate diffeomorphism
invariance~\cite{Albrecht,Barrow}, for which there is therefore
never a question of a varying $c$ being a coordinate artifact.

\section{The speed of light and Maxwell's equations}

As pointed out by Ellis, any VSL theory purporting to represent a
varying speed of the photon should contain a derivation of this
feature from Maxwell's equations, or whatever equations replace
them.

It is, however, the case that in most VSL constructions Maxwell's
theory is derivative, the core being in the proposal of a new
structure, typically dealing with the fate of Lorentz and
diffeomorphism invariance (see the last two sections). Once this
structure is defined, it is usually trivial to set up the matter
action (for all spins, namely 1). The situation has parallels with
setting up Maxwell's theory in General Relativity, where once
minimal coupling is introduced there is not much to say about
Maxwell's theory except that ``derivatives become covariant
derivatives''. There are exceptions to this rule, however. In some
early VSL theories, such as those of Drummond, Hathrell and
Shore~\cite{Drummond2,Shore} changes to the Maxwell action are in
fact at the core of VSL, without any extra structure being
introduced. Here non-minimal couplings to the metric leave the speed
of the photon polarization dependent, a physical prediction that is
derived from first principles.

However, this is not always the case. Take for example bimetric VSL
theories. Once we define the double metric structure, Maxwell's
action is just the same as usual, referred to the matter metric (as
opposed to gravity's metric.) The usual derivations of the photon
phase and group speeds can be made. The standard Maxwell
electromagnetic field action is kept, while $c$ emerges as a
variable physical quantity, when referred to the speed of the
graviton.

This is recognized in~\cite{Ellis} with regards to bimetric
theories, but it is also true in all other theories, for example
in~\cite{Magueijo2}. Here a $x^0$ coordinate is used, but it is
not to be converted into time in the usual way, as explained
in~\cite{Magueijo2}. Again the standard derivations follow, but
when the speed of the photon is identified a new element crucial
to the VSL theory makes this speed variable. The new structure
derives precisely from consideration of the issues of the last 3
sections, namely the concept of minimally coupled clock as the
simplest unit of time for that particular theory. Another
possibility is that the Lorentz symmetry of Maxwell's action is
spontaneously violated or violated by some form of dynamical
symmetry breaking.

Other times, still, the new fundamental structure embodying VSL
requires that Maxwell's action is modified. For example a field
theory realization of DSR necessarily changes the action, either by
invoking higher order derivatives or a non-commutative space-time.
The speed of the photon is then color dependent, something that is
derived from first principles from the new action. The new action,
however, is not the fundamental element of the new theory; rather it
follows from a deformation of the Lorentz group, in this case based
on the choice of a non-linear representation
(see~\cite{Smolin,Smolin2,Kimberly,Dsrfield}).

Once again generalization is not possible; the issue raised must be
addressed within the framework of each theory proposed.

\section{The speed of light and dynamical equations}

More generally Ellis correctly points out that in view of the fact
that $c$ enters into many modern equations of physics, one needs
to consider the effects of a varying $c$ on the whole of physics.
It is not advisable to simply insert a varying speed of light,
$c=c(x)$, into equations before constructing a consistent
dynamical theory of VSL.

But like with Maxwell's theory, this is usually dealt with by the
definition of the new structure embodying a varying $c$ and
explaining the fate of diffeomorphism and Lorentz invariance. For
example, in bimetric
models~\cite{Clayton,Clayton2,Clayton3,Clayton4} there is never any
ambiguity in what to do with any component of the matter action,
electromagnetic or otherwise. Simply write the theory, as before,
making sure you use the matter rather than the gravity metric. The
same applies to the Einstein action, this time derived from a Ricci
scalar made from the gravity metric. A derivation of Einstein's
equations from first principles can then be achieved.

This is equally true in theories of the type defined
in~\cite{Magueijo2}, where all one needs to do is make sure the
connection between $x^0$ and $t$ is suitably modified.
Implications for all sorts of laws (including second quantization)
were discussed in~\cite{Magueijo2}. In Einstein's field equations:
\begin{equation}
G_{\mu\nu}=\kappa T_{\mu\nu}+\Lambda g_{\mu\nu},
\end{equation}
where $\kappa=8\pi G/c^4$ a varying $c$ could equally be
interpreted as a varying $G$ with $c$ kept constant, either choice
leading to different physical conclusions. This is explained
carefully in~\cite{Magueijo2}, where the issue of minimal coupling
to the matter action pegs down which system of units one should be
using, and therefore whether we have a varying $G$ or varying $c$.
The implications for the standard tests of general relativity were
discussed in~\cite{stars}.

The issue raised is indeed a very important one, as in many cases
a varying $c$ does not appear in all kinds of laws.   Take for
example the $c$ that occurs in the fine structure constant,
$\alpha=e^2/\hbar c$. This does not vary in bimetric theories but
does so in theories of the type defined in~\cite{Magueijo2}.

Finally, according to our long experience with constructing
consistent dynamical theories, it is as argued by Ellis advisable to
construct a Lagrangian-Hamiltonian formulation of the theory at the
outset and admit only those solutions to the equations following the
application of a least action principle. This was the route followed
in early VSL papers~\cite{Moffat,Moffat2} and in bimetric
theories~\cite{Clayton,Clayton2,Clayton3,Clayton4,Clayton5,Visser,Drummond}.
But it is not true that a theory has to be defined by a Lagrangian
or a Hamiltonian.

Indeed for centuries theories were defined by the equations of
motion, the Lagrangian formulation being regarded as a
mathematical accessory. We now know that this is far from true.
The Lagrangian formulation brings to the front the concept of
symmetry and how it relates to conservation laws which are
otherwise miraculous accidents. However we must not lose sight of
the fact that experiment is at the root of physics more than
formalism. At the heart of some VSL
theories~\cite{Albrecht,wheeler} is the idea that the laws of
physics may intrinsically evolve. Usually this is precluded by the
fact that we could always define an invariant super-law explaining
how the laws are changing. But not if we state that the time
translational invariance of physics is broken as a result of a
time variability in the speed of light.

This is far from metaphysical and has a very concrete implication:
energy conservation is violated, and nothing, like the proposal of a
new energy form, can be done to fix it. This is in fact the
ingredient behind the solution of the flatness problem proposed
in~\cite{Albrecht}. The Lagrangian formulation may be useful in
bringing this to the front but in many cases it just so happens that
it becomes very awkward. If energy is not conserved, then the
Lagrangian formulation may not be the best way to set up the theory
at all (this is a situation reminiscent of the description of
friction forces).

Absence of a Lagrangian formulation is far from being a general
feature of VSL, but we argue that it may be {\it the} point of those
that attack the philosophical foundations of physics at its most
fundamental level, introducing the concept of intrinsic evolution in
the laws of physics.

\section{Conclusions}

We are glad that reference~\cite{Ellis} gave us the opportunity to
clarify these important foundational issues. However we should
finish by stressing that the success or lack thereof of VSL
theories will likely depend on a variety of other issues.

For VSL theories describing early universe cosmology, it is
natural to compare them to the widely accepted paradigm of
inflation~\cite{Guth,Steinhardt,Linde,Linde2}. Already in an early
paper~\cite{Moffat} and in a more recent work~\cite{Moffat4}, it
was shown that a calculation of the spectrum of primordial
fluctuations can predict successfully a scale invariant Gaussian
spectrum. Such a successful calculation was also performed in a
bimetric model~\cite{Clayton5}. In these calculations the problem
of causally connecting the fluctuations outside the horizon was
achieved through the superluminal speed of light well beyond the
horizon. This superluminal behavior of the speed of light performs
the same role as the superluminal expansion of spacetime in
inflation models.

We are clearly unable to directly observe whether the universe
underwent an inflationary expansion or whether the speed of light
was very large during a short period of time in the early
universe, due to the opaqueness of the surface of last scattering.
However, indirect consequences such as the predicted ratio of the
tensor to scalar modes of the fluctuation spectrum may differ in
VSL and inflationary model calculations, as well as the
predictions for the spectrum of gravitational waves.

On the other hand, the VSL models can be criticized in the same
way as many inflationary models, for fundamental theories of VSL
and inflation are lacking. The question of whether these models
truly solve the initial value problems of the Big Bang model is
still debateable. Both VSL and inflationary models must contend
with the problem of the Big Bang singularity at $t=0$ and possible
violations of the second law of thermodynamics. Although the
possibility of a violation of Lorentz symmetry is now widely
accepted by the physics community, there is as yet no measurable
indication that this actually happens in nature. Such a violation
of Lorentz invariance and possibly diffeomorphism invariance is
part of the VSL paradigm, although it may only occur in the very
early universe.

One reason why VSL may be ahead of inflation is that, at least in
some of its guises, it may be directly testable (and here it is once
again important not to generalize; the following does {\it not}
apply to bimetric and DSR theories). It looks as if the
observational evidence for a redshift dependence in the fine
structure parameter $\alpha$ is here to stay~\cite{webb,sri,webb1}.
No such direct prediction graces the inflationary literature. Of
course a varying alpha may be due to a variety of theories, namely,
the more conservative varying $e$ theories~\cite{bek,haav}. But
there are always distinct predictions differentiating these
theories~\cite{eorc}. We feel that the future of VSL may be not in
its confrontation with inflation but rather in the following:
\begin{itemize}
\item The execution of further high redshift spectroscopic
observations, placing the
Webb results on an even firmer footing.
\item Atomic clock experiments, capable of direct detection
of varying $\alpha$ over the space of a year, as suggested
in~\cite{Magueijo}.
\item An array of supplementary experiments, required for distinguishing
between VSL and other varying $\alpha$ theories (e.g.~\cite{eorc}).
\end{itemize}

Herein may lie the future of VSL.

{\bf Acknowledgements:} We thank George Ellis for providing us with
the motivation to write this paper. The Perimeter Institute is
supported in part by the Government of Canada through NSERC and by
the Province of Ontario through MEDT.

\label{lastpage}


\begin{thebibliography}{99}

\bibitem{Ellis} G. F. R. Ellis, astro-ph/070375.

\bibitem{Uzan} G. F. R. Ellis and J-P Uzan, Amer. J. Phys. {\bf
73}, 240 (2005), gr-qc/0305099.

\bibitem{Magueijo} J. Magueijo, Rept. Prog. Phys. {\bf 66} 2025,
astro-ph/0305457.

\bibitem{Moffat} J. W. Moffat, Int. J. Mod. Phys. {\bf D2},351 (1993),
gr-qc/9211020.


\bibitem{Moffat2} J. W. Moffat, Found. of Phys. {\bf 23}, 411 (1993),
gr-qc/9209001.

\bibitem{Magueijo2} J. Magueijo, Phys. Rev. {D62} 103521 (2000),
gr-qc/0007036.


\bibitem{Albrecht} A. Albrecht and J. Magueijo, Phys. Rev. {\bf
D59}, 043416 (1999), astro-ph/9811018.


\bibitem{Barrow} J. D. Barrow, Phys. Rev. {\bf D59}, 043515
(1999).

\bibitem{Clayton} M. A. Clayton and J. W. Moffat, Phys. Lett.
{\bf B460}, 263 (1999), astro-ph/9812481.

\bibitem{Clayton2} M. A. Clayton and J. W. Moffat, Phys. Lett.
{\bf B477}, 269 (2000), gr-qc/9910112.

\bibitem{Clayton3} M. A. Clayton and J. W. Moffat, Phys. Lett.
{\bf B506}, 177 (2001), gr-qc/0101126.

\bibitem{Clayton4} M. A. Clayton and J. W. Moffat, Int. J. Mod. Phys. {\bf
D11}, 187 (2002), gr-qc/0003070.

\bibitem{Clayton5} M. A. Clayton and J. W. Moffat, JCAP 0307 004 (2003), gr-qc/0304058.

\bibitem{Drummond} I. T. Drummond, Phys. Rev. {\bf D63} 043503
(2001), astro-ph/0008234.

\bibitem{Visser} B. A. Bassett, S. Liberati, C. Molinari-Paris
and M. Visser, Phys. Rev. {\bf D62} 103518 (2000),
astro-ph/0001441.

\bibitem{Smolin} J. Magueijo and L. Smolin, Phys.Rev.Lett. 88: 190403, 2002;
J. Magueijo and L. Smolin, Phys. Rev. D67: 044017, 2003.

\bibitem{Smolin2} J. Magueijo and L. Smolin, Class. Quant. Grav.
{\bf 21}, 1725 (2004), gr-qc/0305055.

\bibitem{Kimberly} D. Kimberly, J. Magueijo, J. Medeiros, Phys.Rev.D70:
084007, 2004.

\bibitem{Drummond2} I. T. Drummond and S. J. Hathrell, Phys. Rev.
{\bf D22}, 343 (1980).

\bibitem{Shore} G. M. Shore, Ann. Phys. (N.Y.) {\bf 128}, 376
(1980).

\bibitem{Moffat3} J. W. Moffat, hep-th/0208122.

\bibitem{Moffat4} J. W. Moffat, Int. J. Mod. Phys. {\bf D12}, 1279 (2003), hep-th/0211167.

\bibitem{Dirac} P. A. M. Dirac, Proc. Roy. Soc. {\bf A209}, 291
(1951).

\bibitem{Bekenstein} J. D. Bekenstein, Phys. Rev. {\bf D66},
123514 (2002), gr-qc/0208081.

\bibitem{Barrow2} J. D. Barrow and J. Magueijo, Phys. Lett. {\bf
B443}, 104 (1998), astro-ph/9811072.

\bibitem{stars}J. Magueijo, Phys.Rev. D63: 043502, 2001.

\bibitem{wheeler}J. Magueijo, ``A genuinely evolving Universe'', in
``Science and ultimate reality'', eds. J. Barrow, P. Davies and
C. Harper, CUP, 2004.

\bibitem{Guth} A. H. Guth, Phys. Rev. {\bf D23}, 347 (1981).

\bibitem{Steinhardt} A. Albrecht and P. Steinhardt, Phys. Rev.
Lett. {\bf 48}1220 (1982).

\bibitem{Linde} A. D. Linde, A. D. Linde, Physz. Lett. {\bf B129},
177 (1983).

\bibitem{Linde2} A. D. Linde, {\it Particle Physics and
Inflationary Cosmology}, (Harwood, Chur, Switzerland, 1990.

\bibitem{bek}J. Bekenstein, Phys.Rev. D25: 1527-1539, 1982

\bibitem{haav}H. Sandvik, J. D. Barrow and J. Magueijo,
Phys.Rev.Lett. 88: 031302, 2002.

\bibitem{eorc}J. Magueijo, J. Barrow and H. Sandvik, Phys.Lett. B549:
284-289, 2002.

\bibitem{Dsrfield}J. Magueijo, Phys.Rev.D73: 124020, 2006.

\bibitem{webb}J. Webb et al, Phys.Rev.Lett. 82 (1999) 884-887;
Phys.Rev.Lett. 87 (2001) 091301.

\bibitem{sri}R. Srianand et al, Phys. Rev. Lett, 92, 121302, 2004;
 R. Chand et al, Astron. Astrophys. 417, 853, 2004.

\bibitem{webb1}M. Murphy, J. Webb and V. Flambaum,
astro-ph/0612407.

\end{thebibliography}
\end{document}